\documentclass [aps,prl,a4paper,reprint,superscriptaddress] {revtex4-1}
\usepackage{amsmath}
\usepackage{graphicx}
\usepackage{subfigure}
\usepackage{xcolor}

\begin{document}
\title{Breath figures under electrowetting: electrically controlled evolution of drop condensation patterns}

\author{Davood Baratian} 
\thanks{Davood Baratian and Ranabir Dey contributed equally.}
\affiliation{Physics of Complex Fluids, MESA+ Institute for Nanotechnology, University of Twente, PO Box 217, 7500 AE Enschede, The Netherlands}

\author{Ranabir Dey} 
\thanks{Davood Baratian and Ranabir Dey contributed equally.}
\affiliation{Physics of Complex Fluids, MESA+ Institute for Nanotechnology, University of Twente, PO Box 217, 7500 AE Enschede, The Netherlands}

\author{Harmen Hoek} 
\affiliation{Physics of Complex Fluids, MESA+ Institute for Nanotechnology, University of Twente, PO Box 217, 7500 AE Enschede, The Netherlands}

\author{Dirk van den Ende}
\affiliation{Physics of Complex Fluids, MESA+ Institute for Nanotechnology, University of Twente, PO Box 217, 7500 AE Enschede, The Netherlands}

\author{Frieder Mugele}
\email[e-mail:]{f.mugele@utwente.nl}
\affiliation{Physics of Complex Fluids, MESA+ Institute for Nanotechnology, University of Twente, PO Box 217, 7500 AE Enschede, The Netherlands}

\begin{abstract}
We show that electrowetting (EW) with structured electrodes significantly modifies the distribution of drops condensing onto flat hydrophobic surfaces by aligning the drops and by enhancing coalescence. Numerical calculations demonstrate that drop alignment and coalescence are governed by the drop size-dependent electrostatic energy landscape that is imposed by the electrode pattern and the applied voltage. Such EW-controlled migration and coalescence of condensate drops significantly alter the statistical characteristics of the ensemble of droplets. The evolution of the drop size distribution displays self-similar characteristics that significantly deviate from classical breath figures on homogeneous surfaces once the electrically-induced coalescence cascades set in beyond a certain critical drop size. The resulting reduced surface coverage, coupled with earlier drop shedding under EW, enhances the net heat transfer.
\end{abstract}
\maketitle
Dropwise condensation of water vapour is intrinsic to natural phenomena like dew formation \cite{beysens1995formation}, and dew/fog harvesting by animals (e.g. Namib Desert beetle) and plants such as Namib Desert plant \cite{malik2014nature}. Dropwise condensation of vapour is also utilized in various technologies like water-harvesting systems \cite{milani2011evaluation}, heat exchangers for cooling systems \cite{kim2002air}, and desalination systems \cite{khawaji2008advances}. The efficacy of these technologies depends on the nucleation, coalescence and growth of the condensate droplets on surfaces, and on their subsequent shedding \cite{beysens2006dew,rose1967mechanism}. The pattern formed by condensing droplets is classically referred to as a breath figure due to its similarity with the pattern formed by breathing on a cold surface \cite{beysens1986growth, viovy1988scaling,fritter1991experiments}. An intriguing feature of breath figures is that the pattern evolution of condensate droplets is self-similar in time, as established by scaling the droplet size distribution \cite{family1988scaling,family1989kinetics,meakin1992droplet,narhe2001difference,blaschke2012breath}. To improve the efficiency of the dropwise condensation process it is essential to control the underlying breath figure characteristics.
 \begin{figure} [ht]
 	\includegraphics*[width=\columnwidth]{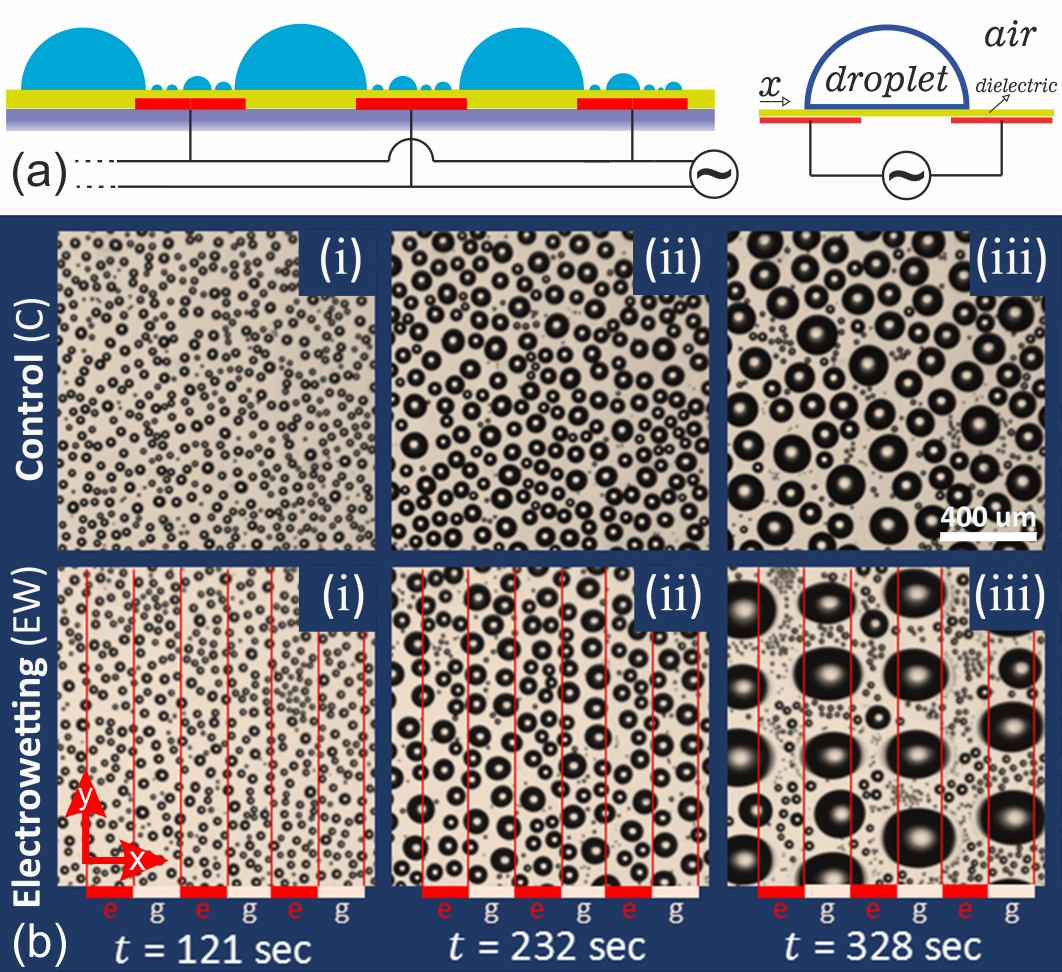}
 	\caption{\label{Fig. 1} (a) Schematic of the substrate used for the condensation experiments. Transparent interdigitated ITO electrodes (red) are patterned on the glass substrate (grey), which is then coated with a hydrophobic dielectric polymer film (green). A schematic of a condensate droplet under EW is also shown. (b) Comparison between breath figures without EW (control) (C-i to C-iii) and under EW ($U_{rms}=150$ V; $f=1$ kHz) (EW-i to EW-iii) at different time $(t)$ instants. The (e)lectrode-(g)ap geometry underneath the dielectric film is indicated by the solid red and white lines. \textcolor{blue}{Gravity points from top-to-bottom i.e. along the negative y--direction.}}
 \end{figure} 

In general, dropwise condensation can be controlled by regulating the mobility of the droplets on the surface. Enhanced dropwise condensation has been studied on superhydrophobic nanostructured surfaces \cite{boreyko2009self,miljkovic2012effect,miljkovic2012jumping,miljkovic2013electric}, on superhydrophobic microgrooved and wettability-patterned surfaces \cite{narhe2004nucleation,ghosh2014enhancing}, and on liquid impregnated textured surfaces \cite{anand2012enhanced,tsuchiya2017liquid}. However, in all the studies so far droplet mobility is altered passively by tailoring the characteristics of the condensing surface through chemical or topographical patterning. In this letter, we go beyond such passive approaches and use alternating (AC) electric field in an electrowetting (EW) configuration with patterned electrodes \cite{mugele2005electrowetting} to actively control the breath figure evolution on otherwise homogeneous hydrophobic surfaces. We demonstrate that EW significantly modifies the distribution of condensate drops using electrical forces, and alters the statistical characteristics of the entire ensemble of droplets from those established for classical breath figures with randomly distributed drops. The breath figure evolution under EW is characterized by size-dependent alignment of the condensate drops at the minima in the corresponding electrostatic energy landscapes and enhanced coalescence.  A scaling analysis shows that the size-distribution of drops is self-similar in time with different characteristics in the initial and final growth phases. The unique transformation in the scaling plot is triggered by EW-induced coalescence cascades beyond a critical drop size. Such EW-mediated alignment and enhanced coalescence increases the average droplet radius, but reduces the surface coverage. 
      
The condensation experiments are performed on a glass plate coated by a hydrophobic dielectric polymer film (Fig. \ref{Fig. 1}(a)). The glass substrate contains a stripe pattern of transparent interdigitated ITO electrodes (Fig. \ref{Fig. 1}(a)). The width of both the electrodes and the gaps is 200 $\mu$m. An AC voltage with a frequency of $f=1$ kHz and a maximum amplitude of 150 V $U_{rms}$ is applied. For the experimental details see S1 in the Supplemental Material \cite{suppmaterial}. For the experiments, a stream of vapour-air mixture at a flow rate of 2 lt/min and a temperature of $36 \pm 1$ $^\circ$C is passed through a condensation chamber, in which the substrate is kept at a temperature of $25$ $^\circ$C. Condensation experiments are performed both without EW (control) and under EW on identical substrates and under identical experimental conditions. The condensation on the control and EW-functionalized surfaces is monitored using a high resolution camera. We denote the time instant at which reliably detectable condensate drops of radius $\sim 5$ $\mu$m are detected for the first time as $t=0$ s (for the image analysis procedure see S8 in \cite{suppmaterial}).    

At the beginning, small drops appear at random locations on both the control and the EW-functionalized surfaces and grow without coalescence (Fig. \ref{Fig. 1}(b)(C-i), (EW-i)). As the drops grow and coalesce frequently  (see Movie S1 in \cite{suppmaterial}), the drops on the EW-functionalized surface initially align parallel to the electrode edges, displaced towards the gap-centres (Fig. \ref{Fig. 1}(b)(EW-ii)). As the droplets grow further and exceed a critical size, sequences of rapid coalescence events (coalescence cascades) create approximately monodisperse drops bridging the gap $(w_g)$ between two adjacent electrodes, and aligned along the gap-centres (Fig. \ref{Fig. 1}(b)(EW-iii); Fig. S2 in \cite{suppmaterial}). At the same time, the drops on the control surface remain randomly distributed, and display more polydispersity and smaller average sizes (Fig. \ref{Fig. 1}(b)(C-iii)). The transformation in the breath figure under EW (from Fig. \ref{Fig. 1}(b)(EW-ii) to (EW-iii)) occurs within a narrow transition period. Eventually, the gravity-induced drop shedding occurs earlier under EW compared to the control surface (Fig. S3 in \cite{suppmaterial}).

To quantify the unique distribution of droplets under EW, we project the droplets in the breath figure onto unit cells of width equal to the pitch $(l)$ of the electrode pattern, and ranging from one electrode-centre to the adjacent electrode-centre. Within a unit cell, we calculate the variation of the droplet area fraction $(\bar{S}_x)$ along the lateral ($x$--) direction. Here, $\bar{S}_x=\sum_{i}A_i/A_{uc}$, where $A_i$ is the $i^{th}$ droplet contact area along a line parallel to the electrode edges ($y$-- direction), and $A_{uc}$ is the unit cell area; $\langle\bar{S}_{\bar{x}}\rangle$ represents the average over all unit cells in the breath figure, and $\bar{x}=x/l$.     
  \begin{figure}[t]
  	\includegraphics*[width=\columnwidth]{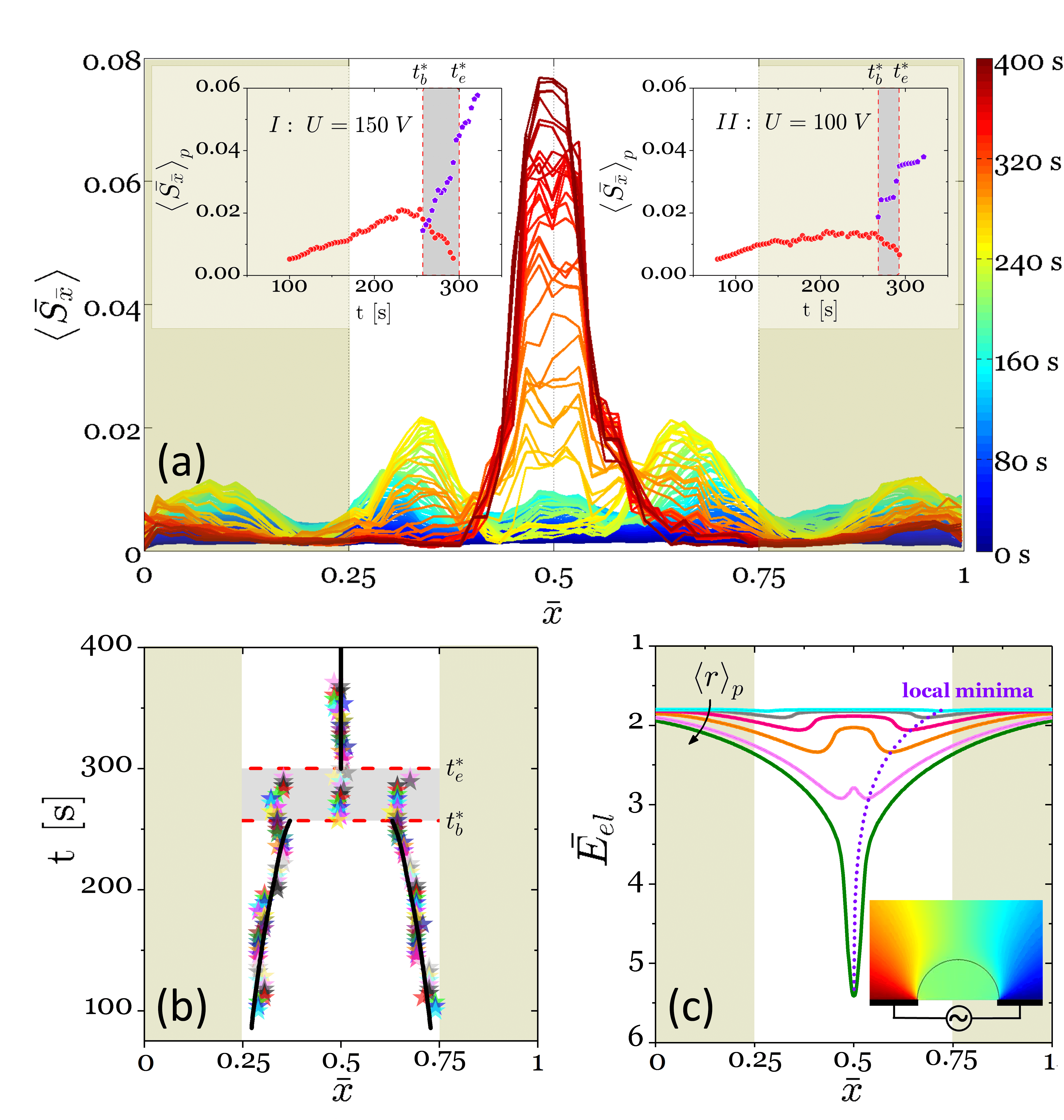}
  	\caption{\label{Fig. 2} (a) Colour-coded temporal evolution of the average area fraction distribution of drops $(\langle\bar{S}_{\bar{x}}\rangle)$ along the non-dimensionalized lateral co-ordinate ($\bar{x}=x/l$), over one pitch of the electrode pattern under EW ($U_{rms}=150$ V). The temporal variations of the peak values $(\langle\bar{S}_{\bar{x}}\rangle_p)$  in $\langle\bar{S}_{\bar{x}}\rangle$ for different voltages are shown in insets-- (I) $U_{rms}=150$ V, (II) $U_{rms}=100$ V. The temporal variations of the peak locations in $\langle\bar{S}_{\bar{x}}\rangle$ are shown by star markers in (b). The transition period is shown in the insets in (a), and in (b), by the grey area. (c) The non-dimensionalized electrostatic energy $(\bar{E}_{el}) $ landscapes corresponding to representative values of the area-weighted average radius $(\langle r \rangle_p=20, 40, 60, 80, 100, 110 \; \mu m)$ of the drops constituting the peak(s) in $\langle\bar{S}_{\bar{x}}\rangle$ before (with two minima) and after transition (with a single minimum). The evolution in the electrostatic energy minima locations with increasing $\langle r \rangle_p$ is shown by the solid black line in (b). Inset shows the electric potential $(\varphi)$ distribution for a condensate droplet under EW (schematic in Fig. \ref{Fig. 1}(a)).} 
  \end{figure}
The temporal evolution of $\langle\bar{S}_{\bar{x}}\rangle$ shows the gradual development of two similar peaks from the gap edges (Fig. \ref{Fig. 2}(a)). The development of these local maxima describes the alignment of the condensate drops on both sides of the gap-centre (Fig. \ref{Fig. 1}(b)(EW-ii)). In contrast, the corresponding $\langle \bar{S}_{\bar{x}} \rangle$ distribution on the control surface remains uniform (Fig. S4 in \cite{suppmaterial}). At any time instant, the peak value $(\langle\bar{S}_{\bar{x}}\rangle_p)$ describes the maximum droplet coverage. Over time, as the droplets coalesce and grow (see Movie S1), $\langle\bar{S}_{\bar{x}}\rangle_p$ gradually increases (red symbols in inset (I) in Fig. \ref{Fig. 2}(a)), and concurrently, the locations of the two peaks gradually shift towards the gap-centre (symbols in Fig. \ref{Fig. 2}(b)). As the average radius of the drops constituting a peak $(\langle r\rangle_p)$ exceeds a critical value $\langle r\rangle^*_{p}\sim 0.3 w_g$, the coalescence cascades set in. Consequently, a new peak in $\langle\bar{S}_{\bar{x}}\rangle$ emerges at the gap-centre (purple symbols in inset (I) in Fig. \ref{Fig. 2}(a)) while the side peaks start to decay; this marks the beginning  $(t^*_b)$ of the transition period. Within the transition period, the side peaks disappear and the peak at the gap-centre survives in the end $(t^*_e)$ (Fig. \ref{Fig. 2}(a) and inset (I), and Fig. \ref{Fig. 2}(b)). This single peak in $\langle\bar{S}_{\bar{x}}\rangle$ reflects the alignment of the bigger condensate drops at the gap-centre (Fig. \ref{Fig. 1}(b)(EW-iii)). Beyond $t^*_e$, these drops grow (Fig. \ref{Fig. 2}(a) and inset (I)) while remaining aligned along the gap-centre (Fig. \ref{Fig. 2}(b)). A similar breath figure evolution is observed for other values of the applied voltage, as shown by the $\langle \bar{S}_{\bar{x}} \rangle_p$ evolution at $U_{rms}=100$ V (inset (II) in Fig. \ref{Fig. 2}(a)).

\begin{figure}[t]
  	\includegraphics*[width=\columnwidth]{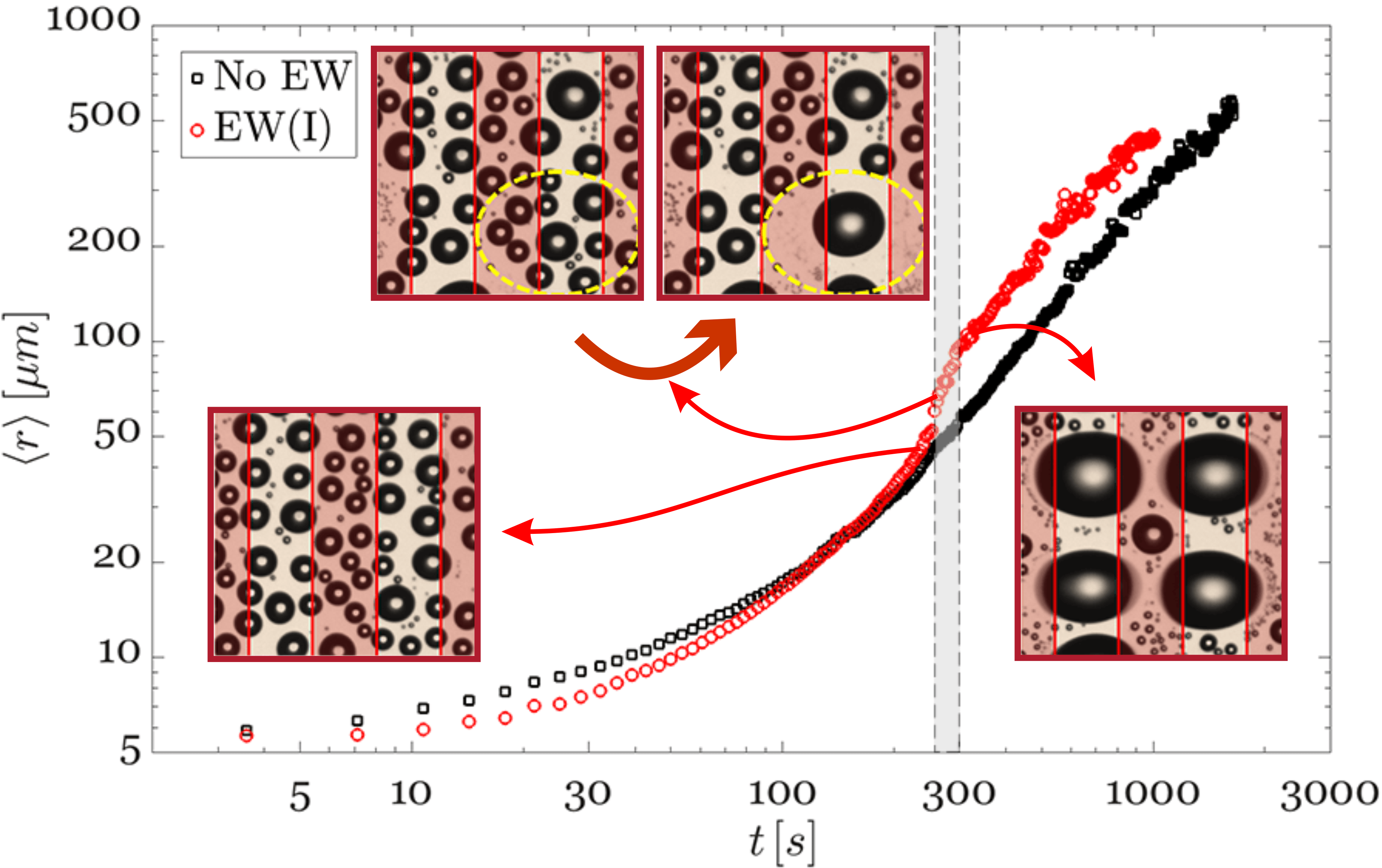}
  	\caption{\label{Fig. 4} Evolutions of the area-weighted mean radius $(\langle r \rangle)$ of the breath figure droplets without EW (control) and under EW ($U_{rms}=150$ V) till the first shedding. The droplet patterns corresponding to different growth regimes of $\langle r \rangle$ under EW are shown as insets. The similar $\langle r \rangle$ variation for $U_{rms}=100$ V is shown in Fig. S6 in \cite{suppmaterial}.} 
  \end{figure}

To understand the evolution of $\langle\bar{S}_{\bar{x}}\rangle$ under EW, we calculate the electrostatic energy $(E_{el})$ profile over the unit cell for a representative condensate droplet of variable size (schematic in Fig. \ref{Fig. 1}(a)). For a droplet having radius $r=\langle r\rangle_p$, the electric potential ($\varphi$) corresponding to a $x-$position is calculated by solving $\nabla \cdot [(\epsilon_0 \epsilon -i\frac{\sigma}{\omega})\nabla \varphi] =0$ using a finite element method (inset in Fig. \ref{Fig. 2}(c)). Here, $\epsilon_0$ is the vacuum permittivity, $\epsilon$ is the material dielectric constant, $\sigma$ is the material electrical conductivity, and $\omega$ is the circular frequency. For the present experimental conditions, the contribution of polarization current is significantly small compared to the conductive contribution. Hence, the system operates in the AC-EW regime rather than `dielectrowetting' \cite{mchale2011dielectrowetting} (see S9 in \cite{suppmaterial} for numerical computation details). Thereafter, the total electrostatic energy of the system is calculated as $E_{el}=-\int_v \frac{1}{2} \vec{E} \cdot \epsilon_0 \epsilon \vec{E} \; dv$, where $\vec{E}=-\nabla \varphi$ is the electric field, and $v$ is the computation domain volume. Subsequently, the $E_{el}(x)$ landscape (non-dimensionalized as $\bar{E}_{el}=|E_{el}|/4 \pi \langle r\rangle_p^2 \gamma$; $\gamma$ is the water-air surface tension) is evaluated as a function of the drop position along the unit cell. Before the onset of coalescence cascade i.e. for $\langle r\rangle_p \lesssim \langle r\rangle^*_p$, $\bar{E}_{el}$ is symmetric about the gap-centre with minima (electrostatic potential wells) on either side of it (Fig. \ref{Fig. 2}(c)). The locations of these two potential wells gradually shift towards the gap-centre with increasing $\langle r\rangle_p$ (dotted line in Fig. \ref{Fig. 2}(c)). The temporal evolution of the peak locations in $\langle\bar{S}_{\bar{x}}\rangle$ (symbols) closely follow the electrostatic energy minima locations (solid black line) corresponding to increasing $\langle r\rangle_p$ (Fig. \ref{Fig. 2}(b)). The drops formed due to coalescence under EW thus migrate to the electrostatic energy minima corresponding to their sizes, culminating in the alignment of the drops on either side of the gap-centre. As $\langle r\rangle_p$ exceeds $\langle r\rangle^*_p \sim 0.3 w_g$, droplets under the side peaks with $r>\langle r\rangle^*_p$ migrate towards the corresponding electrostatic energy minima even closer to the gap-centre. Motion of these droplets initiates collisions between droplets on both sides of the gap-centre triggering the coalescence cascades (S5 in \cite{suppmaterial}). The resulting drops with diameters comparable to $w_g$ $(r\gtrsim 0.6w_g)$, along with the adjacent electrodes, form two parallel plate capacitors in series, with the dielectric layer as spacers \cite{dieter2014trapping}. In this case, $\bar{E}_{el}$ is symmetric with a single minimum at the gap-centre (Fig. \ref{Fig. 2}(c)). Consequently, these approximately monodisperse drops migrate to, and remain aligned along, the gap-centre (Fig. \ref{Fig. 2}(b)). Such alignment results in a definite periodicity of the droplet pattern along the $x-$ direction which is determined by the electrode pitch (Fig. \ref{Fig. 1}(b)(EW-iii)). Furthermore, EW-induced coalescence cascades result in a sharp increase of the area-weighted average radius $(\langle r \rangle=\sum r^3/\sum r^2)$ of the droplets (Fig. \ref{Fig. 4}). The larger droplets created due to the coalescence cascades increase $\langle r \rangle$, compared to that for the control surface (Fig. \ref{Fig. 4}). \textcolor{blue}{Eventually, the gravity-driven drop shedding under AC-EW sets in at a smaller drop size as compared to the control surface (Fig. \ref{Fig. 4}), due to the reduction of effective contact angle hysteresis under AC-EW \cite{li2008make} and the consequential enhancement of gravity-driven drop mobilization \cite{t2011electrically}.}
  \begin{figure}[t]
  	\includegraphics*[width=\columnwidth]{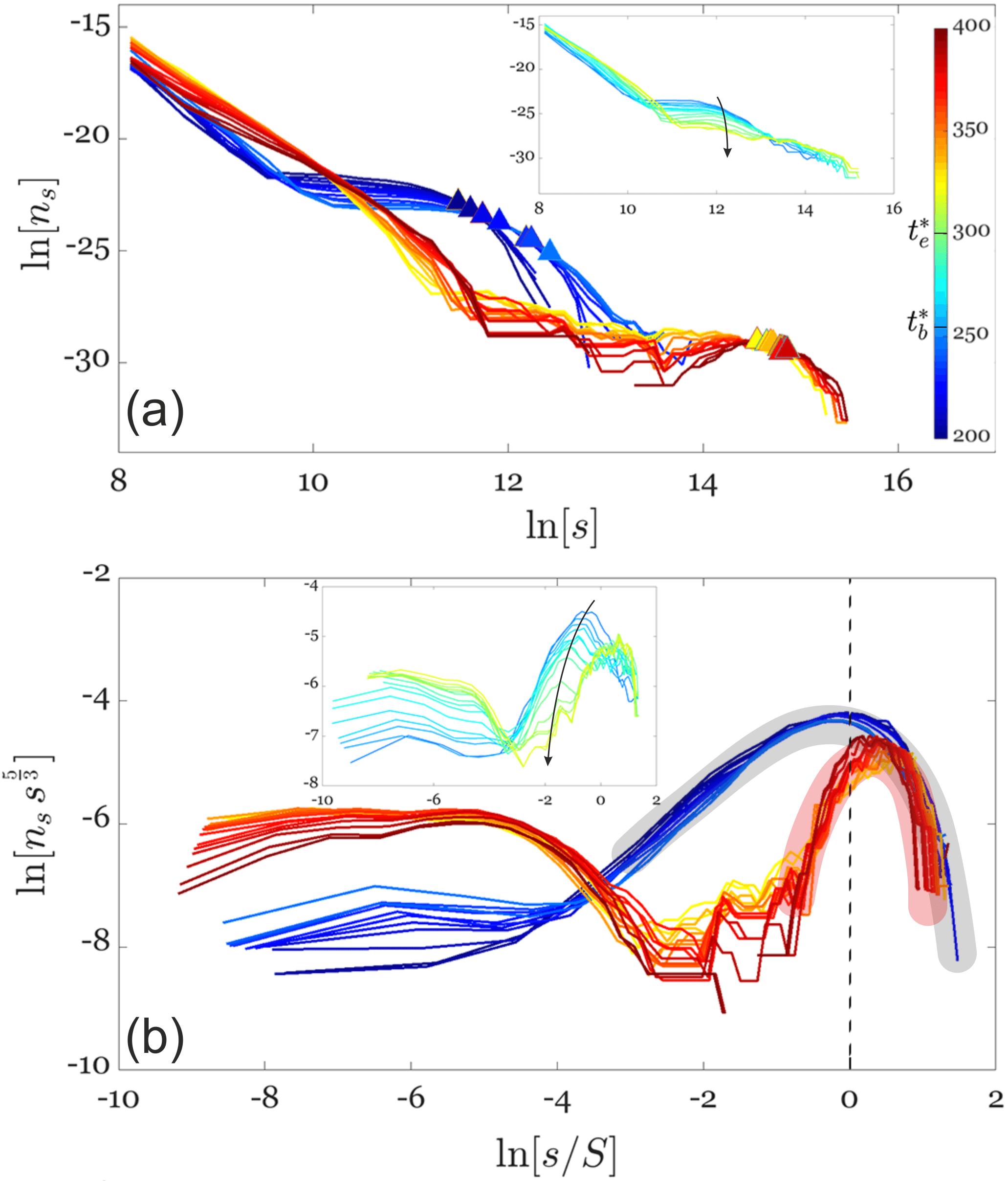}
  	\caption{\label{Fig. 3} (a) Self-similar evolution of the droplet size $(s)$ distribution $(n_s)$ in the breath figure under EW  ($U_{rms}=150$ V) before and after the transition period. The time evolution is colour coded with times before and after transition coloured in shades of blue and red respectively. The triangular markers show the $n_s$ estimates for average drop sizes constituting the peaks in $\langle\bar{S}_{\bar{x}}\rangle$ (Fig.\ref{Fig. 2}(a)). (b) Corresponding scaling plots of $n_s$ obtained using Eq. \ref{Eq.1} before and after the transition. The $n_s$ evolution within the transition period, and the corresponding scaling plots, are shown in the insets in (a) and (b) respectively.} 
  \end{figure}      
  \begin{figure}[t]
  	\includegraphics*[width=\columnwidth]{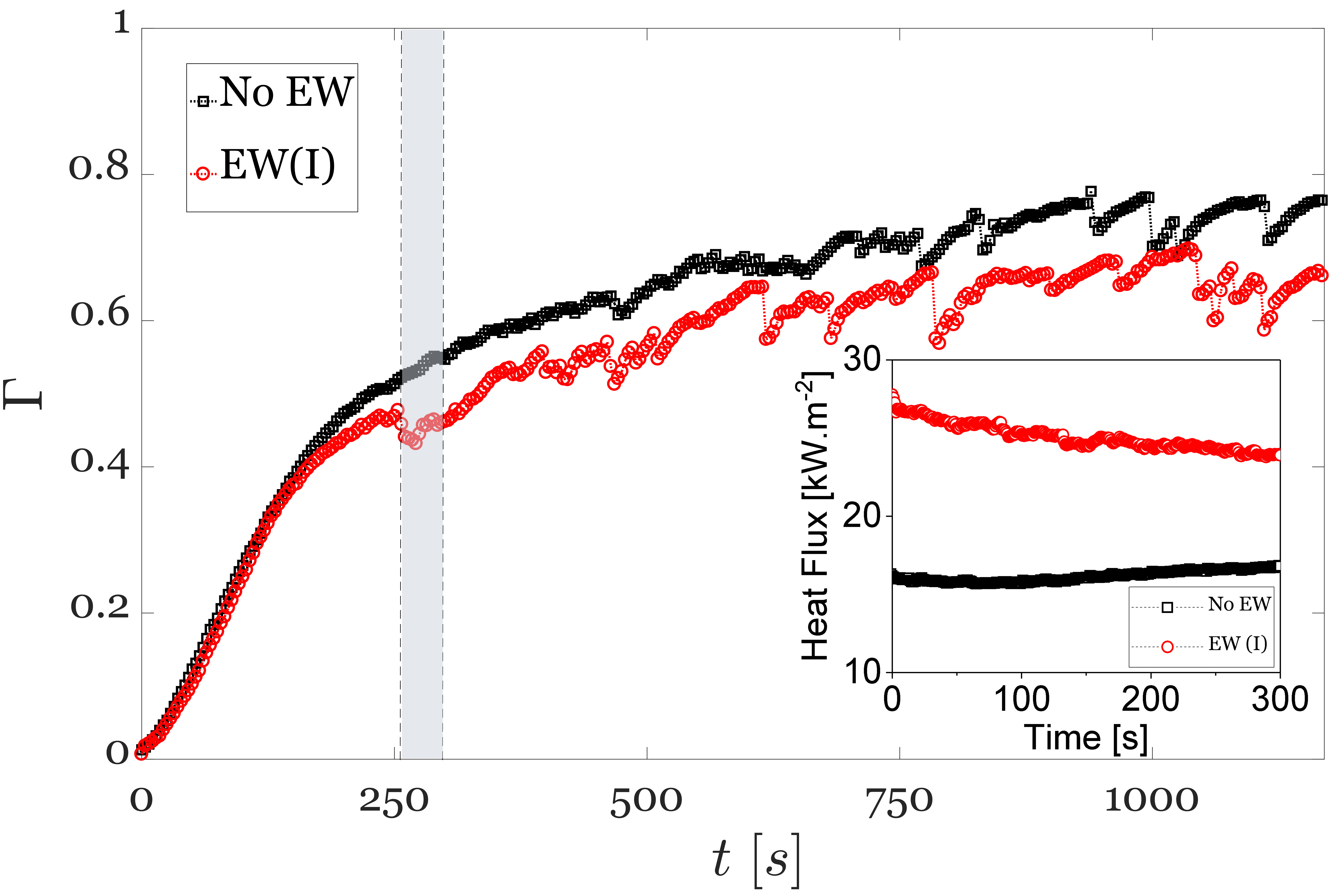}
  	\caption{\label{Fig. 5} Evolutions of the breath figure surface coverage $(\Gamma)$ without EW (control) and under EW ($U_{rms}=150$ V). Inset shows the enhanced heat flux under EW. Heat transfer data are recorded in a separate setup; see S10 in \cite{suppmaterial} for details.} 
  \end{figure}   
  
The EW-mediated evolution of the droplet size distribution is also significantly different from that established for classical breath figures. The droplet size $(s \sim r^3)$ distribution under EW is bimodal exhibiting a power-law decay for smaller droplets and a bell-shaped distribution for larger droplets (Fig. \ref{Fig. 3}(a)). Here, $n_s(s,t)$ represents the number of droplets of volume $s$ per unit droplet volume and surface area. The temporal variation of $n_s$ is self-similar obeying the scaling relation \cite{family1988scaling,family1989kinetics,meakin1992droplet,barenblatt1996scaling}
  \begin{equation}
  n_s(t)=s^{-\varTheta}f(\frac{s}{S}) \label{Eq.1}
  \end{equation}       
(Fig. \ref{Fig. 3}(b) and Fig. S7 in \cite{suppmaterial}). Here, $S(t) \sim \langle r \rangle^3$, and the exponent $\varTheta=1+d/D$ depends on the dimensionality of the drops $(D=3)$ and the condensing surface  $(d=2)$. For $t<t_b^*$, the coalescence-dominated evolution of the bell-shaped distributions for the larger drops is collapsed by the similarity transformation (Eq. \ref{Eq.1}) on considering $\varTheta=5/3$ (grey shaded region of the scaling plots for $t<t_b^*$ in Fig. \ref{Fig. 3}(b)). It must be noted that the smaller drops $(ln[s/S]<-4)$ follow a different scaling that is dependent on the nucleation and the small scale growth mechanisms \cite{blaschke2012breath, barenblatt1996scaling}. However, for characterizing the breath figures under EW (Fig. \ref{Fig. 2}(a)), it is sufficient to describe the evolution of the bell-shaped part of the droplet size distribution, since the corresponding droplets align within the gaps. This is substantiated by the fact that the $n_s$ estimates corresponding to $s_p \sim (\langle r \rangle_p)^3$ are close to the peak of the bell-shaped distribution (triangular markers in Fig \ref{Fig. 3}(a)). Naturally, the evolution of $n_s(s,t)$ under EW loses its self-similarity in the transition period due to the EW-induced coalescence cascades (inset in Fig. \ref{Fig. 3}(b)), unlike the droplet size distribution for the control, which remains self-similar throughout the coalescence-dominated growth regime (Fig. S7 in \cite{suppmaterial}). However, for $t\gtrsim t^*_e$, the evolution of  $n_s$ for the larger droplets again exhibits self-similarity, albeit with a significantly different functional form compared to that for $t<t_b^*$ (Fig. \ref{Fig. 3}(b)) and for the control(Fig. S7 in \cite{suppmaterial}). The almost uniform sizes of the large droplets created due to the coalescence cascades manifest in the narrower bell-shaped distribution for $t \gtrsim t^*_e$ (compare the blue versus red scaling plots in Fig. \ref{Fig. 3}(b)). Furthermore, due to the large separation of sizes between the dominant monodisperse droplets and the small background droplets (Fig. \ref{Fig. 1}(b)(EW-iii)), the corresponding bell-shaped distributions are no longer peaked at $S$, contrary to that for  $t<t_b^*$ and for the control surface (Fig. \ref{Fig. 3}(b); Fig. S7 in \cite{suppmaterial}). Interestingly, the alignment of the growing monodisperse drops at the electrostatic energy minima, which always remain at the gap-centre, makes the corresponding size distribution analogous to that for condensation on a line $(d=1)$ \cite{family1989kinetics}. Such evolution of the droplet size distribution is unique to the EW-induced anisotropy in the system. Furthermore, the significant reduction in $n_s$ for the relatively larger droplets, due to the coalescence cascades (Fig. \ref{Fig. 3}(b)), results in enhanced `release' of substrate surface area. Consequently, we observe $\sim10-12 \%$ reduction in the steady state surface coverage ($\Gamma=\sum \pi r^2/A_{FOV}$; $A_{FOV}$ is the area of the field of view) under EW, with respect to the control surface (Fig. \ref{Fig. 5}). \textcolor{blue}{The reduced $\Gamma$ coupled with the enhanced gravity-driven shedding of condensate drops, due to the AC-EW-induced increased droplet growth and reduced hysteresis, results in enhanced heat transfer (inset in Fig. \ref{Fig. 5}). Preliminary heat transfer measurements show a striking $50-60\%$ increase in net heat flux for dropwise condensation under EW (inset in Fig. \ref{Fig. 5}).} It must be noted that thinner inorganic dielectric layers (e.g. SiO$_2$ or Al$_2$O$_3$) are available for EW \cite{berry2006low,yang2017high} to significantly minimize the thermal resistance compared to the presently used polymer layers.

In summary, we have shown that the electrostatic energy landscape under AC-EW induces migration and coalescence of condensate droplets leading to unique breath figure characteristics. The coalescing drops align at the corresponding electrostatic energy minima, instead of staying restricted to the centre of mass of the parent drops as in classical breath figures. The resulting periodicity of the droplet pattern under EW can be further controlled by simply tuning the electrode geometry. We hope that this study will trigger a general theoretical analysis of drop condensation patterns in arbitrary energy landscapes. We have also shown that the modification of the breath figure characteristics under EW leads to enhanced heat transfer. We anticipate that these effects will be very useful for optimizing applications involving dropwise condensation, like heat exchangers and breath figure templated self-assembly \cite{lin2012evaporative}.

\begin{acknowledgments}
We sincerely thank Kripa Varanasi for allowing us to use the heat transfer measurement setup in his lab, and Karim Khalil for help during the heat transfer measurements. We thank Arjen Pit for assistance with the numerical calculations, Daniel Wijnperle and B. Robert for their help with the preparation of the condensation substrates. We acknowledge financial support by the Dutch Technology Foundation STW, which is part of the Netherlands Organization for Scientific Research (NWO), and the VICI program (grant 11380).
\end{acknowledgments}
        
\bibliographystyle{apsrev4-1}
\bibliography{ew_condensation_bf_SEP2017}
\end{document}